


\font\twelverm=cmr10 scaled 1200    \font\twelvei=cmmi10 scaled 1200
\font\twelvesy=cmsy10 scaled 1200   \font\twelveex=cmex10 scaled 1200
\font\twelvebf=cmbx10 scaled 1200   \font\twelvesl=cmsl10 scaled 1200
\font\twelvett=cmtt10 scaled 1200   \font\twelveit=cmti10 scaled 1200

\skewchar\twelvei='177   \skewchar\twelvesy='60


\def\twelvepoint{\normalbaselineskip=12.4pt
  \abovedisplayskip 12.4pt plus 3pt minus 9pt
  \belowdisplayskip 12.4pt plus 3pt minus 9pt
  \abovedisplayshortskip 0pt plus 3pt
  \belowdisplayshortskip 7.2pt plus 3pt minus 4pt
  \smallskipamount=3.6pt plus1.2pt minus1.2pt
  \medskipamount=7.2pt plus2.4pt minus2.4pt
  \bigskipamount=14.4pt plus4.8pt minus4.8pt
  \def\rm{\fam0\twelverm}          \def\it{\fam\itfam\twelveit}%
  \def\sl{\fam\slfam\twelvesl}     \def\bf{\fam\bffam\twelvebf}%
  \def\mit{\fam 1}                 \def\cal{\fam 2}%
  \def\tt{\twelvett}
  \textfont0=\twelverm   \scriptfont0=\tenrm   \scriptscriptfont0=\sevenrm
  \textfont1=\twelvei    \scriptfont1=\teni    \scriptscriptfont1=\seveni
  \textfont2=\twelvesy   \scriptfont2=\tensy   \scriptscriptfont2=\sevensy
  \textfont3=\twelveex   \scriptfont3=\twelveex  \scriptscriptfont3=\twelveex
  \textfont\itfam=\twelveit
  \textfont\slfam=\twelvesl
  \textfont\bffam=\twelvebf \scriptfont\bffam=\tenbf
  \scriptscriptfont\bffam=\sevenbf
  \normalbaselines\rm}



\def\beginlinemode{\endmode
  \begingroup\parskip=0pt \obeylines\def\\{\par}\def\endmode{\par\endgroup}}
\def\beginparmode{\endmode
  \begingroup \def\endmode{\par\endgroup}}
\let\endmode=\par
{\obeylines\gdef\
{}}
\def\singlespace{\baselineskip=\normalbaselineskip}

\def\oneandahalfspace{\baselineskip=\normalbaselineskip
  \multiply\baselineskip by 3 \divide\baselineskip by 2}
\def\doublespace{\baselineskip=\normalbaselineskip \multiply\baselineskip by 2}

\newcount\firstpageno
\firstpageno=2
\footline={\ifnum\pageno<\firstpageno{\hfil}%
\else{\hfil\twelverm\folio\hfil}\fi}
\let\rawfootnote=\footnote              
\def\footnote#1#2{{\rm\singlespace\parindent=0pt\rawfootnote{#1}{#2}}}
\def\raggedcenter{\leftskip=4em plus 12em \rightskip=\leftskip
  \parindent=0pt \parfillskip=0pt \spaceskip=.3333em \xspaceskip=.5em
  \pretolerance=9999 \tolerance=9999
  \hyphenpenalty=9999 \exhyphenpenalty=9999 }
\def\dateline{\rightline{\ifcase\month\or
  January\or February\or March\or April\or May\or June\or
  July\or August\or September\or October\or November\or December\fi
  \space\number\year}}
\def\received{\vskip 3pt plus 0.2fill
 \centerline{\sl (Received\space\ifcase\month\or
  January\or February\or March\or April\or May\or June\or
  July\or August\or September\or October\or November\or December\fi
  \qquad, \number\year)}}


\hsize=6.5truein
\vsize=8.9truein
\parskip=\medskipamount
\twelvepoint            
\oneandahalfspace       
\overfullrule=0pt       



\def\title                      
  {\null\vskip 3pt plus 0.2fill
   \beginlinemode \doublespace \raggedcenter \bf}

\def\author                     
  {\vskip 3pt plus 0.2fill \beginlinemode
   \singlespace \raggedcenter}

\def\affil                      
  {\vskip 3pt plus 0.1fill \beginlinemode
   \oneandahalfspace \raggedcenter \sl}

\def\abstract                   
  {\vskip 3pt plus 0.3fill \beginparmode
   \oneandahalfspace \narrower ABSTRACT: }

\def\endtitlepage               
  {\endpage                     
   \body}

\def\body                       
  {\beginparmode}               

\def\head#1{                    
  \filbreak\vskip 0.5truein     
  {\immediate\write16{#1}
   \raggedcenter \uppercase{#1}\par}
   \nobreak\vskip 0.25truein\nobreak}

\def\subhead#1{                 
  \vskip 0.25truein             
  {\raggedcenter #1 \par}
   \nobreak\vskip 0.25truein\nobreak}

\def\refto#1{$|{#1}$}           

\def\references                 
  {\subhead{References}         
   \beginparmode
   \frenchspacing \parindent=0pt \leftskip=1truecm
   \parskip=8pt plus 3pt \everypar{\hangindent=\parindent}}

\gdef\refis#1{\indent\hbox to 0pt{\hss#1.~}}    

\gdef\journal#1, #2, #3, 1#4#5#6{               
    {\sl #1~}{\bf #2}, #3, (1#4#5#6)}           

\def\refstylenp{                
  \gdef\refto##1{ [##1]}                                
  \gdef\refis##1{\indent\hbox to 0pt{\hss##1)~}}        
  \gdef\journal##1, ##2, ##3, ##4 {                     
     {\sl ##1~}{\bf ##2~}(##3) ##4 }}

\def\refstyleprnp{              
  \gdef\refto##1{ [##1]}                                
  \gdef\refis##1{\indent\hbox to 0pt{\hss##1)~}}        
  \gdef\journal##1, ##2, ##3, 1##4##5##6{               
    {\sl ##1~}{\bf ##2~}(1##4##5##6) ##3}}

\def\pr{\journal Phys. Rev., }

\def\endreferences{\body}

\def\figurecaptions             
  { \beginparmode
   \subhead{Figure Captions}
}

\def\endpage                    
  {\vfill\eject}

\def\endpaper                   
  {\endmode\vfill\supereject}


\def\ref#1{Ref. #1}                     
\def\Ref#1{Ref. #1}                     

\def\frac#1#2{{\textstyle{#1 \over #2}}}
\def\half{{\textstyle{ 1\over 2}}}

\def\etc{{\it etc.}}

\def\sla{\raise.15ex\hbox{$/$}\kern-.57em}
\def\leaderfill{\leaders\hbox to 1em{\hss.\hss}\hfill}
\def\twiddle{\lower.9ex\rlap{$\kern-.1em\scriptstyle\sim$}}
\def\bigtwiddle{\lower1.ex\rlap{$\sim$}}
\def\gtwid{\mathrel{\raise.3ex\hbox{$>$\kern-.75em\lower1ex\hbox{$\sim$}}}}
\def\ltwid{\mathrel{\raise.3ex\hbox{$<$\kern-.75em\lower1ex\hbox{$\sim$}}}}
\def\square{\kern1pt\vbox{\hrule height 1.2pt\hbox{\vrule width 1.2pt\hskip 3pt
   \vbox{\vskip 6pt}\hskip 3pt\vrule width 0.6pt}\hrule height 0.6pt}\kern1pt}

\def\m@th{\mathsurround=0pt }
\def\leftrightarrowfill{$\m@th \mathord\leftarrow \mkern-6mu
 \cleaders\hbox{$\mkern-2mu \mathord- \mkern-2mu$}\hfill
 \mkern-6mu \mathord\rightarrow$}
\def\overleftrightarrow#1{\vbox{\ialign{##\crcr
     \leftrightarrowfill\crcr\noalign{\kern-1pt\nointerlineskip}
     $\hfil\displaystyle{#1}\hfil$\crcr}}}


\font\titlefont=cmr10 scaled\magstep3

\def\martinstyletitle                      
  {\null\vskip 3pt plus 0.2fill
   \beginlinemode \doublespace \raggedcenter \titlefont}

\font\twelvesc=cmcsc10 scaled 1200

\def\author                     
  {\vskip 3pt plus 0.2fill \beginlinemode
   \singlespace \raggedcenter\twelvesc}


\def\heading                            
  {\vskip 0.5truein plus 0.1truein      
   \beginparmode \def\\{\par} \parskip=0pt \singlespace \raggedcenter}

\def\subheading                         
  {\vskip 0.25truein plus 0.1truein     
   \beginlinemode \singlespace \parskip=0pt \def\\{\par}\raggedcenter}

\def\tag#1$${\eqno(#1)$$}

\def\align#1$${\eqalign{#1}$$}

\def\aligntag#1$${\gdef\tag##1\\{&(##1)\cr}\eqalignno{#1\\}$$
  \gdef\tag##1$${\eqno(##1)$$}}

\def\overset #1\to#2{{\mathop{#2}\limits^{#1}}}
\def\underset#1\to#2{{\let\next=#1\mathpalette\undersetpalette#2}}
\def\undersetpalette#1#2{\vtop{\baselineskip0pt
\ialign{$\mathsurround=0pt #1\hfil##\hfil$\crcr#2\crcr\next\crcr}}}


\def\ref#1{Ref.~#1}                     
\def\Ref#1{Ref.~#1}                     
\def\[#1]{[\cite{#1}]}
\def\cite#1{{#1}}
\def\(#1){(\call{#1})}
\def\call#1{{#1}}
\def\taghead#1{}
\def\frac#1#2{{#1 \over #2}}
\def\half{{\frac 12}}

\def\12{{1\over2}}

\def\etc{{\it etc.\ }}

\def\sla{\raise.15ex\hbox{$/$}\kern-.57em}
\def\leaderfill{\leaders\hbox to 1em{\hss.\hss}\hfill}
\def\twiddle{\lower.9ex\rlap{$\kern-.1em\scriptstyle\sim$}}
\def\bigtwiddle{\lower1.ex\rlap{$\sim$}}
\def\gtwid{\mathrel{\raise.3ex\hbox{$>$\kern-.75em\lower1ex\hbox{$\sim$}}}}
\def\ltwid{\mathrel{\raise.3ex\hbox{$<$\kern-.75em\lower1ex\hbox{$\sim$}}}}
\def\square{\kern1pt\vbox{\hrule height 1.2pt\hbox{\vrule width 1.2pt\hskip 3pt
   \vbox{\vskip 6pt}\hskip 3pt\vrule width 0.6pt}\hrule height 0.6pt}\kern1pt}
\def\tdot#1{\mathord{\mathop{#1}\limits^{\kern2pt\ldots}}}

\def\pmb#1{\setbox0=\hbox{#1}%
  \kern-.025em\copy0\kern-\wd0
  \kern  .05em\copy0\kern-\wd0
  \kern-.025em\raise.0433em\box0 }

\catcode`@=11
\newcount\r@fcount \r@fcount=0
\newcount\r@fcurr
\immediate\newwrite\reffile
\newif\ifr@ffile\r@ffilefalse
\def\w@rnwrite#1{\ifr@ffile\immediate\write\reffile{#1}\fi\message{#1}}

\def\writer@f#1>>{}
\def\referencefile{
  \r@ffiletrue\immediate\openout\reffile=\jobname.ref%
  \def\writer@f##1>>{\ifr@ffile\immediate\write\reffile%
    {\noexpand\refis{##1} = \csname r@fnum##1\endcsname = %
     \expandafter\expandafter\expandafter\strip@t\expandafter%
     \meaning\csname r@ftext\csname r@fnum##1\endcsname\endcsname}\fi}%
  \def\strip@t##1>>{}}

\def\citeall#1{\xdef#1##1{#1{\noexpand\cite{##1}}}}
\def\cite#1{\each@rg\citer@nge{#1}}	

\def\each@rg#1#2{{\let\thecsname=#1\expandafter\first@rg#2,\end,}}
\def\first@rg#1,{\thecsname{#1}\apply@rg}	
\def\apply@rg#1,{\ifx\end#1\let\next=\relax
\else,\thecsname{#1}\let\next=\apply@rg\fi\next}

\def\citer@nge#1{\citedor@nge#1-\end-}	
\def\citer@ngeat#1\end-{#1}
\def\citedor@nge#1-#2-{\ifx\end#2\r@featspace#1 
  \else\citel@@p{#1}{#2}\citer@ngeat\fi}	
\def\citel@@p#1#2{\ifnum#1>#2{\errmessage{Reference range #1-#2\space is bad.}%
    \errhelp{If you cite a series of references by the notation M-N, then M and
    N must be integers, and N must be greater than or equal to M.}}\else%
 {\count0=#1\count1=#2\advance\count1 by1\relax\expandafter\r@fcite\the\count0,
  \loop\advance\count0 by1\relax
    \ifnum\count0<\count1,\expandafter\r@fcite\the\count0,%
  \repeat}\fi}

\def\r@featspace#1#2 {\r@fcite#1#2,}	
\def\r@fcite#1,{\ifuncit@d{#1}
    \newr@f{#1}%
    \expandafter\gdef\csname r@ftext\number\r@fcount\endcsname%
                     {\message{Reference #1 to be supplied.}%
                      \writer@f#1>>#1 to be supplied.\par}%
 \fi%
 \csname r@fnum#1\endcsname}
\def\ifuncit@d#1{\expandafter\ifx\csname r@fnum#1\endcsname\relax}%
\def\newr@f#1{\global\advance\r@fcount by1%
    \expandafter\xdef\csname r@fnum#1\endcsname{\number\r@fcount}}

\let\r@fis=\refis			
\def\refis#1#2#3\par{\ifuncit@d{#1}
   \newr@f{#1}%
   \w@rnwrite{Reference #1=\number\r@fcount\space is not cited up to now.}\fi%
  \expandafter\gdef\csname r@ftext\csname r@fnum#1\endcsname\endcsname%
  {\writer@f#1>>#2#3\par}}

\def\ignoreuncited{
   \def\refis##1##2##3\par{\ifuncit@d{##1}%
    \else\expandafter\gdef\csname r@ftext\csname r@fnum##1\endcsname\endcsname%
     {\writer@f##1>>##2##3\par}\fi}}

\def\r@ferr{\endreferences\errmessage{I was expecting to see
\noexpand\endreferences before now;  I have inserted it here.}}
\let\r@ferences=\references
\def\references{\r@ferences\def\endmode{\r@ferr\par\endgroup}}

\let\endr@ferences=\endreferences
\def\endreferences{\r@fcurr=0
  {\loop\ifnum\r@fcurr<\r@fcount
    \advance\r@fcurr by 1\relax\expandafter\r@fis\expandafter{\number\r@fcurr}%
    \csname r@ftext\number\r@fcurr\endcsname%
  \repeat}\gdef\r@ferr{}\endr@ferences}


\let\r@fend=\endpaper\gdef\endpaper{\ifr@ffile
\immediate\write16{Cross References written on []\jobname.REF.}\fi\r@fend}

\catcode`@=12

\citeall\refto		
\citeall\ref		%
\citeall\Ref		%

%
%

\def\refto#1{$^{#1}$}         

\def\la{\langle}
\def\ra{\rangle}

\def\n{{\bar n}}
\def\N{{\bar N}}
\def\x{{\bf x}}

\def\H{{H_T}}
\def\a{\alpha}
\def\b{\beta}
\def\d{{\delta}}
\def\s{{\sigma}}
\def\e{{\epsilon}}
\def\l{{\lambda}}

\def\Tr{{\rm Tr}}
\def\ih{i}
\def\au{{\underline{\alpha}}}

\def\s{{\sigma}}

\def\A{{\cal A}}

\def\P{{\bar P}}
\def\p{{\partial }}
\def\do{{\downarrow}}
\def\up{{\uparrow}}

\centerline{\bf Decoherence of Hydrodynamic Histories: A Simple Spin Model}

\author T.A.~Brun
\affil
Department of Physics
Queen Mary and Westfield College 
University of London
Mile End Road
London E1 4NS
UK
\vskip 0.2in
\centerline{\rm and}
\vskip 0.1in
\author J.J.~Halliwell 
\affil
Theory Group
Blackett Laboratory
Imperial College
London SW7 2BZ
UK
\vskip 0.5in
\centerline {\rm Preprint QMW-PH-95-48. December, 1995}
\vskip 0.1in 
\centerline {\rm Submitted to {\sl Physical Review D}}
\endpage
\vskip 1.0in
\endpage

\abstract
In the context of the decoherent histories approach to the quantum
mechanics of closed systems, Gell-Mann and Hartle have argued that
the variables typically characterizing the quasiclassical domain of
a large complex system are the integrals over small volumes of
locally conserved densities -- hydrodynamic variables. The aim of
this paper is to  exhibit some simple models in which approximate
decoherence arises as a result of local conservation. We derive a
formula which shows the explicit connection between local
conservation and approximate decoherence.   We then consider a class
of models consisting of a large number of weakly interacting
components, in which the projections onto local densities may be
decomposed into projections onto one of two alternatives of the
individual components.  The main example we consider  is a
one-dimensional chain of locally coupled spins, and the projections
are onto the total spin in a subsection of the chain.  We compute
the decoherence functional for histories of local densities, in the
limit when the number of components is very large. We find that
decoherence requires two things: the smearing volumes must be
sufficiently large to ensure approximate conservation, and the local
densities must be partitioned into sufficiently large ranges to
ensure protection against quantum fluctuations.

\endpage

\head{\bf 1. Introduction}

One of the primary aims of quantum cosmology is to understand how
the universe is classical to a very high degree of precision, given
the hypothesis that it is described at the most fundamental level by
quantum theory [\cite{Har1,Har6,Hal4}]. 
Mathematically, this aim
translates into the question of why, on sufficiently large scales,
quantum theory  admits an emergent description of the universe
involving only a small number of dynamical variables obeying an
approximately closed set of deterministic evolution equations. 

Whilst many approaches to this question concentrate on the
positions and momenta of point particles, these dynamical variables
are special cases of a more general description of the physical
world in terms of local densities: number density, momentum density,
energy density, charge density, {\it etc.} The question of emergent
classicality then consists first, of understanding why these
variables enjoy such a distinguished role, and second, of deriving
the familiar hydrodynamic equations for these densities.

The decoherent histories approach to quantum theory is a recently
developed formulation of quantum theory which is particularly suited
to this problem [\cite{GH1,GH2,GH3,Gri,Omn,Omn2}].  In brief, the
approach permits
predictions to be made in genuinely closed systems, such as the
entire universe, without relying on notions of measurement or on an
external classical domain. The goal of the approach is to assign
probabilities to the histories of a closed quantum system. It is
this feature that makes it particularly useful for discussing
emergent classicality. This is because to assert in quantum theory
that a certain variable approximately satisfies a deterministic
evolution equation involves computing the probability for a
time-ordered sequence of values of that variable, {\it i.e.}, for a
history. It is then necessary to show  that the probability for that
history is strongly peaked about the sequence of values
corresponding to the evolution equation. 

This paper constitutes the beginnings of a general application of
the decoherent histories approach to the question of deriving
hydrodynamic equations. 

We begin with a very brief review of the decoherent histories approach
[\cite{GH1,GH2,GH3,Gri,Omn,Omn2,DoH}], followed by a general
discussion of the issue of emergent classicality for hydrodynamic variables.
A more detailed statement of the particular models considered
in this paper may be found at the end of this section.

The models and technical results presented here are admittedly
simple, perhaps to the extent of being overshadowed by the very
broad discussion of emergent classicality given below. This paper
is, however, a small first step in a possibly extensive programme.
We have therefore chosen to give a general sketch of that programme,
to indicate the direction of future research and to
provide the general context in which the possible significance of 
our modest results may be understood.

\subhead{\bf 1(A). The Decoherent Histories Approach to Quantum Theory}

A quantum-mechanical history is a sequence of propositions at a
succession of times. Propositions  at a fixed moment of time  are
represented by sets of projections operators, $\{ P_{\a} \} $. They
are exhaustive and exclusive,
$$\sum_{\a} P_{\a} =1, \quad \quad
P_{\a} P_{\beta} = \delta_{\a \beta} \ P_{\a}
\eqno(1.1)$$
A projector is said to be fine-grained if it is of the form
$ | \a \ra \la \a | $, where $\{| \a \ra \}$ are a complete set of
states. Otherwise it is said to be coarse-grained.
A single quantum-mechanical history is characterized by a string of
time-dependent projections,
$P_{\a_1}^1(t_1), \cdots, P_{\a_n}^n(t_n)$, together with an initial
state $\rho$. The time-dependent projections are related to the
time-independent ones by
$$
P^k_{\a_k}(t_k) = e^{i H(t_k-t_0)} P^k_{\a_k} e^{-i H(t_k-t_0)} 
\eqno(1.2)
$$
where $H$ is the Hamiltonian.
The candidate probability for such histories is
$$
p(\a_1, \a_2, \cdots \a_n) = {\rm Tr} \left( P_{\a_n}^n(t_n)\cdots 
P_{\a_1}^1(t_1)
\rho P_{\a_1}^1 (t_1) \cdots P_{\a_n}^n (t_n) \right)
\eqno(1.3)
$$

To be a true probability, Eq.(1.3) must satisfy the probability sum
rules. That is, it must be such that the probability for each
coarser-grained history is the sum of the probabilities for the
constituent finer-grained histories.  Coarser-grained histories may
be constructed, for example, by summing the projections at each
moment of time: 
$$ 
{\bar P}_{{\bar \a}} = \sum_{\a \in {\bar \a} }
P_{\a} 
\eqno(1.4)
$$
The probability sum rules are generally not satisfied due to
quantum interference.

Sets of histories which do not suffer interference, and hence for
which the sum rules are satisfed may be found using the decoherence functional,
$$
D({\underline {\a}} , {\underline {\a}'} ) = 
\Tr \left( P_{\a_n}^n(t_n)\cdots 
P_{\a_1}^1(t_1)
\rho P_{\a_1'}^1 (t_1) \cdots P_{\a_n'}^n (t_n) \right)
\eqno(1.5)
$$
Here $ {\underline {\a}} $ denotes the string $\a_1, \a_2, \cdots
\a_n$. Intuitively, the decoherence functional measures the amount
of interference between pairs of histories.
It may be shown that
the probability sum rules are satisfied for all
coarse-grainings if and only if 
$$
Re D({\underline {\a}} , {\underline {\a}'} ) = 0 
\eqno(1.6)
$$
for all distinct pairs of
histories ${\underline {\a}}, {\underline {\a}'}$ [\cite{Gri}].
Such sets of histories are said to be {\it consistent} (or 
{\it weakly decoherent}). 

The consistency condition (1.6) is usually (but not always)
satisfied only for coarse-grained histories.
When sets of histories satisfy the consistency
condition (1.6) as a result of coarse-graining, they often
satisfy, in addition, the stronger condition of
{\it decoherence},
$$
D(\au, \au') = 0, \quad for \quad \au \ne \au'
\eqno(1.7)
$$
(sometimes classified as {\it medium} 
decoherence, with yet more stringent criteria for {\it strong} 
decoherence [\cite{GH3}]).
Physically, decoherence  is intimately related to the existence of
records about the system somewhere in the universe [\cite{GH2}].

In most cases of interest decoherence is only approximate,
so measures of approximate decoherence are required. We say that
the degree of decoherence is of order $\e$ if the probability sum
rules are violated only up to terms of order $\e$ times the
probabilities themselves [\cite{DoH}]. It may be shown that, under certain
conditions on the distribution of off-diagonal terms in the
decoherence functional, decoherence to order $\e$ will be satisfied
if
$$
\bigl| D (\au, \au') \bigr|^2 \ < \ \e^2 \ D(\au, \au) D(\au', \au').
\eqno(1.8)
$$

\subhead {\bf 1(B). Emergent Quasiclassicality}

We would like to use the decoherent
histories approach  to demonstrate the emergence of an approximately
classical world from an underlying quantum one, together with the
quantum fluctuations about it described by the standard Copenhagen
quantum mechanics of measured subsystems. Such a state of affairs is
referred to as a quasiclassical domain [\cite{GH1,GH2,GH3}]. 
In more technical terms, a quasiclassical domain consists of a
decoherent set of histories, characterized largely by the same types
of variables at different times, and whose probabilities are peaked
about deterministic evolution equations for the variables
characterizing the histories. 

The histories should, moreover, be {\it maximally refined} with
respect to a specified degree of approximate decoherence.  That is,
one specifies a decoherence factor $\epsilon$ in the approximate
decoherence condition discussed above. This should, for example, be
chosen so that the probabilities are defined to a precision far
beyond any conceivable test. Then, the histories should be
fine-grained ({\it e.g.}, by reducing the widths of the projections)
to the point that further fine-graining would lead to violation of
the specified degree of approximate decoherence. The resulting set
of histories are then called maximally refined. The reason for
maximally refining the histories is to reduce as much as possible
any apparent subjective element in the choice of coarse-graining.

Given the Hamiltonian and initial state of the system, one's task is
to compute the decoherence functional for various different choices
of histories, and see which ones lead to quasiclassical behaviour.
We expect this to be a formidable task, and there is no reason to
believe that there will be a unique answer. Generally, one might
expect that there will be a hierarchy of variables, similar to the
BBGKY hierarchy [\cite{CaH}].

Many previous discussions [\cite{GH2,DoH,PZ,Zur,JoZ}]
of emergent classicality concern systems
in which there is a natural separation of the total closed sytem
into ``system'' and ``environment'', and this separation is the
source of the coarse-graining required for 
decoherence.\footnote{$^{\dag}$}{More precisely, by separation
into system and environment, we mean that the total Hilbert space
for the closed system may be written as a tensor product of the
system and environment Hilbert spaces. Some authors appear to
use this expression to mean something more general.} A generic closed
system, however,  will usually not have such a separation, and it is
one of the strengths of the decoherent histories approach that it
does not rely on the existence of one. Certain variables will,
however, be distinguished by the existence of conservation laws for
total energy, momentum, charge, particle number, {\it etc.}
Associated with such conservation laws are local conservation laws
of the form
$$
{\partial \rho \over \partial t} +{\bf  \nabla} \cdot {\bf j} = 0
\eqno(1.9)
$$
The candidate quasiclassical variables are then
$$
Q_V = \int_{V} d^3 x \ \rho({\bf x})
\eqno(1.10)
$$
These are the local densities discussed above.
If the volume $V$ becomes 
infinite, $Q_V$ will be an exactly conserved quantity. In quantum
mechanics it will commute with the Hamiltonian, and, as is easily 
seen, histories of $Q_V$'s will decohere exactly [\cite{HLM}]

If the volume is finite but large compared to
the microscopic scale, $Q_V$ will be slowly varying compared to 
other dynamical variables. This is because the local conservation
law (1.9) permits $Q_V$ to change only by redistribution, which
is limited by the rate at which the locally conserved quantity can
flow out of the volume. Because these quantities are slowly varying,
histories of them should approximately decohere.

More precisely, when the volume $V$ is infinite in size, the
decoherence functional for histories of the variables (1.10) will be
exactly diagonal. As $V$ shrinks from infinite size, the decoherence
functional will develop off-diagonal terms. However, one
would not expect the off-diagonal terms to grow appreciably until
$V$ approaches the length scales characteristic of the system at
hand. These length scales can depend only on the initial state and
on the Hamiltonian. The length scale associated with the initial
state could in principle take any value, hence  some restriction on
the initial state will be necessary. The length scales associated
with Hamiltonians of the type we are
typically interested in will generally be very small. The various
physically relavent scales, are, for example, the mean free path
between molecular collisions, or the length scale of intermolecular
forces. Therefore, for some class of initial states (to be
determined), one would expect the off-diagonal terms of the
decoherence functional to remain small until $V$ shrinks down
to the microscopic scale.

Hence, for a suitable class of initial states,
we expect the variables (1.10) to be approximately decoherent as a
consequence of their association with conservation laws, which are
in turn connected with the global symmetries of the system.

Given decoherence, we may then examine the probabilities for
histories of hydrodynamic variables, and ask whether they are peaked
about the expected hydrodynamic equations.  Derivations of the
hydrodynamics equations from an underlying microscopic quantum
theory have certainly been carried out before (see Ref.[\cite{For}]
for example).  These derivations have shown that the {\it
expectation values} of the local densities $Q_V$, in a local
equilibrium state, evolve in approximate accordance with the
hydrodynamic equations. The derivation  contemplated here, however,
is considerably more general (although undoubtedly related in some
way). In the decoherent histories approach we would like to show
that the probabilities for histories of imprecisely specified values
of the local densities, for a more general class of initial states,
are peaked about the hydrodynamic equations.  This paper is only a
first step in that direction; but it is important to keep the ultimate
goal in mind.

The connection between the earlier derivations of the hydrodynamic
equations and the histories derivation contemplated here is 
analagous to the connection between the density operator and the
decoherent histories approaches to emergent classicality of quantum
Brownian motion models. In the density operator approach, it was
shown that an initial Gaussian wave packet followed an approximately
classical path, with dissipation, and with fluctuations due to
classical and quantum noise [\cite{Zur,PZ,HaZ}]. In the decoherent histories
approach, it was shown that for a wide variety of initial states,
the probabilities for histories of position samplings is strongly
peaked about classical evolution, with dissipation, with a width of
peaking depending on classical and quantum noise [\cite{GH2,DoH}].

\subhead{\bf 1(C). This Paper}

The programme sketched above is clearly a very extensive one.
Although clear in principle how to proceed, it is very difficult to
carry out in practice. In principle, since the operators are known,
one may compute their spectrum, from which the projectors onto
ranges of the spectrum may be derived. The time evolution of those
projectors may be deduced and inserted in the decoherence
functional. In practice, this is very difficult and it seems likely
that new mathematical techniques will be needed. 

It is unlikely, for example, that the path integral techniques  so
successful in the study of quantum Brownian motion models will be
effective here  [\cite{GH2,DoH}]. In the quantum
Brownian motion models, the interesting variables are the position
or momentum of a distinguished particle, and projections onto these
variables are easily implemented as restrictions on the paths in a
sum over paths. The hydrodyanmic variables considered here, however,
are non-trivial functions of positions and momenta, and projections
onto ranges of their spectra cannot in general be expressed in terms
of restrictions in a  sum over paths in phase space.

The quantum Brownian motion models also made heavy use of the
influence functional method [\cite{FeV,CaL}]. Again this is
inapplicable, because it relies on being able to explicitly 
integrate out the ``environment'' ({\it i.e.}, the variables ignored
in the coarse-graining procedure), which cannot be done here. 

To make a start on the general problem what is required is some very
simple models which retain the essential features of approximate
decoherence through approximate conservation, yet are simple enough
to be solvable. In this paper we will present some models which
are of this type.

In Section II, we derive a formula expressing the general connection
between approximate conservation and approximate decoherence.
We will not use this formula explicitly in this paper, although
we present it to flesh out the heuristic arguments given above, and
because it may be of practical use elsewhere.

In Section III, we consider a system consisting of a large number
$N$ of
weakly interacting components in which the individual components are
described by just two alternatives. An example is a large number of
non-interacting particles in a box divided into two sections, and
the two alternatives for each individual component are that the
particle is in the left or right section of the box. A crude
hydrodynamic variable is then the number of particles in, say,
the right hand section. The projection onto such a hydrodynamic
variable can be given in terms of the projections onto the
individual particles, and we derive a formula which gives 
this connection precisely. Another example is the spin system model we
describe below. For this class of models, we compute the decoherence
functional for histories characterized by projections onto densities
of large collections of particles. We thus compute the degree of
decoherence.

The calculation of Section III is principally concerned with the
combinatoric aspects of large collections of particles. The feature
of decoherence through approximate conservation, with which we are
ultimate concerned, is to be found in the decoherence functionals
for the individual components. Hence we need to compute such a
decoherence functional explicitly. This is carried out for a spin
model in Section IV.

The model consists of a long chain of $M$ locally coupled spins, and
the Hamiltonian conserves the total spin. We employ the simplest
coarse-graining possible, which is to divide the chain into two
pieces, of lengths $M_1$ and $M_2$ (so $M_1 + M_2 = M $), and project
onto the amount of spin into one section of the chain.

The ground state of the system is the state in which all the spins
point up. The first excited states, the so-called spin waves,
consist of superpositions of particle-like states in which one of
the spins is down. The higher excited states are more difficult to
construct because they involve interactions between spin waves, but
they may be approximated by suitable combinations of the first
excited states in the limit that the coupling is weak. In this
approximation the system is therefore not unlike a collection of
weakly interacting particles. We thus show that the decoherence
functional for projections onto the number of, say, down spins in
one section may be approximated by a product of decoherence
functionals for the ``one particle'' states, in which the
projections ask in which side is the down spin.  This means that, in
this  approximation, our spin model is a system of the type
considered in Section III.

We calculate the decoherence functional for the individual
components numerically. It is approximately diagonal, but not
surprisingly, the degree of decoherence is not very good, being no
better than would be expected from the overlap of a random pair of
states in a large Hilbert space. Its significant feature is the way
the degree of decoherence changes  as $M_1$ and $M_2$ are changed.
We use this result, together with the formula derived in Section
III, to compute the degree of decoherence for projections onto 
ranges of spin density.

We summarize and conclude in Section V.

\head{\bf 2. Approximate Conservation and Approximate Decoherence}

In this section we briefly outline the connection between
approximate conservation and approximate decoherence.

As in Section I, let $Q$ denote a local density averaged over a
volume $V$:
$$
Q = \int_V  d^3 x \ \rho (\x)
\eqno(2.1)
$$
By integrating the conservation equation for $\rho $, Eq.(1.9),
over a time
interval $[0,t]$ and over the spatial volume $V$, it is readily
shown that
$$
Q_t = Q - \int_0^t dt' \int_{\partial V} d^2 x \ {\bf n} \cdot {\bf
j} (\x, t')
\eqno(2.2)
$$
where $\partial V$ is the boundary of $V$.
This relation will also hold in the 
quantum theory for suitably ordered operators.

Now consider the decoherence functional for histories characterized
by projections onto these operators at two moments of time:
$$
D(\a_1, \a_2 | \a_1', \a_2 )
= \Tr \left( P_{\a_2} e^{ - \ih H t} P_{\a_1} \rho P_{\a_1'} e^{ \ih
H t} \right)
\eqno(2.3)
$$
This may be written
$$
D(\a_1, \a_2 | \a_1', \a_2 )
= \sum_{m,n} 
\la m | P_{\a_2} e^{ - \ih H t} P_{\a_1} | n \ra
\ \rho_{nm} \la m | P_{\a_2} e^{ - \ih H t} P_{\a_1'} | n \ra^*
\eqno(2.4)
$$
where $ |n \ra $ denotes a complete set of states, and $\rho_{nm}
= \la n | \rho | m \ra $. The degree of decoherence will therefore
depend on the size of the amplitudes,
$$
\A_{mn} (\a_1, \a_2) = 
\la m | P_{\a_2} e^{ - \ih H t} P_{\a_1} | n \ra 
\eqno(2.5)
$$
Projections onto exactly conserved quantities will commute
with $H$, and $\A_{mn} (\a_1, \a_2)$ will be exactly zero unless
$ \a_2 = \a_1 $, hence the decoherence functional will be exactly
diagonal. We are interested, however, in the case in which
the projections commute only approximately with the
Hamiltonian.

Suppose the spectrum of the operator $Q$ is continuous. Then
projections onto
precisely specified values of its eigenvalues may be constructed
using the delta-function:
$$
P_{\a} = \d ( Q - \a )
\eqno(2.6)
$$
(These are not fine-grained projections, since the
eigenvalues will typically be highly degenerate). Projections onto
imprecisely specified ranges of eigenvalues are readily constructed
by summing over ranges of $\a$, although we will not do this
explicitly here.

The amplitudes $\A_{mn} (\a_1, \a_2)$ may now be written
$$
\eqalignno{
\A_{mn} (\a_1, \a_2) &= \la m | \d (Q- \a_2 ) \ e^{ \ih H t} \ \d ( Q - \a_1 ) | n
\ra
\cr
& = \la m | \d (Q- \a_2 ) \ e^{ - i \e Q } \ e^{ i \e Q} \ e^{ \ih H t}
\ \d ( Q - \a_1 ) |
n \ra
&(2.7)
\cr }
$$
where $\e$ is an arbitrary parameter. Now we use the fact that
$$
e^{ i \e Q_t} = e^{ \ih H t} \ e^{ i \e Q} \ e^{ - \ih H t}
\eqno(2.8)
$$
where $Q_t = e^{ \ih H t} Q e^{ - \ih H t} $.
Hence we have
$$
\A_{mn} (\a_1, \a_2) =
\la m | \d (Q- \a_2 ) \ e^{ - i \e Q } \ e^{- \ih H t}
\ e^{ i \e Q_t} \ \d ( Q - \a_1 ) | n \ra
\eqno(2.9)
$$
$Q_t$ is also given by (2.2), so inserting (2.2), expanding to first
order in $\e$, using the fact that $ Q \delta (Q -\a ) =
\a \delta (Q - \a ) $, and rearranging, we get
$$
\A_{mn} (\a_1, \a_2) = - { 1 \over (\a_2 - \a_1) } \int_0^t dt' 
\int_{\partial V}
d^2 x \ \la m | \d ( Q - \a_2 ) \ e^{ \ih H t} \ {\bf n} \cdot
{\bf j} (\x, t') \ \d (Q- \a_1 ) | n \ra
\eqno(2.10)
$$
Eq.(2.10) is the main result of this section. Inserted in the
decoherence functional Eq.(2.4), (2.10) yields the degree of
decoherence as a function of the size of the smearing volume $V$. In
particular, we see that decoherence becomes exact as the boundary
$\p V $ of $V$ goes to infinity, as it must, since the operators $Q$
are then exactly conserved. Of course, the rate at which approximate
decoherence approaches exact decoherence will depend on  the states
$ | n \ra $, $ | m \ra $, in accordance with our general
expectations. 

In the case of exact conservation
the amplitude $\A_{mn} (\a_1, \a_2)$ is non-zero only for
$ \a_2 = \a_1 $, hence the probability of $\a_2$ given
$\a_1 $ is $1$ when $ \a_2 = \a_1$, and zero otherwise.
In the case of approximate conservation, however, one expects the
situation to be a little more complicated. Given decoherence,
the probability distribution for histories will not necessarily
be peaked about $\a_2 = \a_1 $. Rather, since we expect the
probabilities for histories to indicate hydrodynamic equations,
it will be peaked about values of $\a_1, \a_2 $ consistent
with some deterministic evolution equations. For this reason,
it is perhaps better to speak of approximate
determinism, rather than approximate conservation.

\head{\bf 3. Decoherence of Densities of Large Collections of
Non-Interacting Particles}

\subhead{\bf 3(A). The System}

Consider a system which consists of a very large number $N$ of
particles or components whose interactions are so weak that they may
be neglected. We are interested in the
case in which each component is described by just two alternatives
at each moment of time, which may be represented by projections
$ P $ and $ \P = 1 - P $. 
An example is a collection of particles in a
box divided into two sections, and the projections $P$ and $\P$ then
represent the propositions that the particle is in, respectively,
the right-hand or left-hand section of the box. A version of this
system involving spins will be described in the next section.
We will refer to the alternatives represented by $P$ and $\P$ as
``yes'' and ``no'', respectively.

Using these elementary projections onto the individual components of
the system, projections onto densities of the whole system may
be constructed. For a system of two particles, for example, the
number of particles in the
right-hand section of the box, may be 2, 1 or 0. These
propositions are represented, respectively, by the projections,
$$
\eqalignno{
P_2 &= P \otimes P 
&(3.1)
\cr
P_1 &= P \otimes \P + \P \otimes P
&(3.2)
\cr
P_0 &= \P \otimes \P
&(3.3)
\cr }
$$
It is easy to see that these projections are mutually exclusive and
exhaustive, as required.

For large systems, the projections onto densities rapidly
become quite cumbersome. 
However, the
following trick turns out to be extremely useful. 
We have the identity,
$$
{ 1 \over 2 \pi} \int_{- \pi}^{\pi} d \l \ e^{ - i \l n} e^{  i
\l m} = \d_{nm}
\eqno(3.4)
$$
where $ \d_{nm} $ is the Kronecker delta. The
projection operator onto number density $\n$ in a system of $N$
components is then given by
$$
P_{\n} = {1 \over 2 \pi} \int_{- \pi}^{\pi} d \l
\ e^{- i \l \n} 
\ F_1 (\l) \otimes F_2 (\l) \cdots \otimes F_N (\l)
\eqno(3.5)
$$
Here
$$
F_k (\l) = e^{i \l} P^{(k)} + \P^{(k)}
\eqno(3.6)
$$
where $P^{(k)}$ and $\P^{(k)}$ denote the projections $P$, $\P$
operating on component number $k$. What happens in Eq.(3.5) is that
in the tensor product over the $F_k(\l)$'s, the projection onto
number density $n$ occurs with coefficient, $ e^{i \l n} $, {\it
i.e.},
$$
\eqalignno{
\ F_1 (\l) \otimes F_2 (\l) \cdots \otimes F_N (\l)
=& \P \otimes \P \otimes \cdots \otimes \P
\cr
& + e^{  i \l} \left( P \otimes \P \otimes \cdots \otimes \P
+ \P \otimes P \otimes \cdots \otimes \P + \cdots\right)
\cr
& + \cdots
\cr
=& \sum_{n=0}^N \ e^{ i \l n} \ P_n
&(3.7)
\cr }
$$
The integration over $\l$ then picks out only the term with
coefficent $e^{i \l \n}$.

These projections are onto precise values of number density.
Generally one is interested in imprecise values of number density,
{\it i.e.}, whether the number density lies in a specified range,
and these projections are obtained by summing over $\n$. We will 
consider this in more detail below.

\subhead{\bf 3(B). The Decoherence Functional}

We may now write down the decoherence functional for histories
characterized by projections at two moments of time onto precise
values of number density. It is,
$$
\eqalignno{
D & (\n_1, \n_2 | \n_1', \n_2 ) =
\Tr \left( P_{\n_2} e^{ - \ih Ht } P_{\n_1} \rho
P_{\n_1'} e^{ \ih H t} P_{\n_2} \right)
\cr
&= { 1 \over (2 \pi)^4 } 
\int_{-\pi}^{ \pi} d \l_1
\int_{-\pi}^{ \pi} d \l_1'
\int_{-\pi}^{ \pi} d \l_2
\int_{-\pi}^{ \pi} d \l_2'
\ e^{ -i \l_1 \n_1 +  i \l_1' \n_1' 
-  i \l_2 \n_2 + i \l_2' \n_2 }
\cr
& \times
\Tr \left( F_1 (\l_2) \otimes F_2 (\l_2 ) \cdots \otimes F_N (\l_2)
\ e^{- \ih \H t} \ F_1 (\l_1) \otimes F_2 (\l_1) \cdots \otimes F_N (\l_1)
\right.
\cr
& \ \times \left.
\ \rho_T \ F^{\dag}_1 (\l_1') \otimes F^{\dag}_2 (\l_1' ) 
\cdots \otimes F^{\dag}_N (\l_1') 
\ e^{ \ih \H t}
\ F^{\dag}_1 (\l_2') \otimes F^{\dag}_2 (\l_2' ) 
\cdots \otimes F^{\dag}_N (\l_2') \right)
&(3.8)
\cr}
$$
Note that we can of course use the simplifying feature
$ \left( P_{\n_2} \right)^2 = P_{\n_2} $ in the final projection,
but it will become clear below why we have not done this.

To make further progress we make two simplifications. First, we
assume that the interaction between component particles is
negligible, so the Hamiltonian $\H$ for the total system has the form
$$
\H = H \otimes 1 \cdots \otimes 1 + 1 \otimes H \cdots \otimes 1
+ \cdots
\eqno(3.9)
$$
Second, we assume that the intial density operator $\rho_T$ for the
total system factors, and that the density operator for each
component system is the same:
$$
\rho_T = \rho \otimes \rho \cdots \otimes \rho
\eqno(3.10)
$$
The decoherence functional may now be written
$$
\eqalignno{
D(\n_1, \n_2 | \n_1', \n_2 ) 
&= { 1 \over (2 \pi)^4 } 
\int_{-\pi}^{ \pi} d \l_1
\int_{-\pi}^{ \pi} d \l_1'
\int_{-\pi}^{ \pi} d \l_2
\int_{-\pi}^{ \pi} d \l_2'
\ e^{ - i \l_1 \n_1 +  i \l_1' \n_1' 
-  i \l_2 \n_2 +  i \l_2' \n_2 }
\cr
& \times \prod_{k=1}^N \ \Tr \left(
F_k (\l_2 ) \ e^{ -\ih H t } \ F_k (\l_1) \ \rho
\ F^{\dag}_k (\l_1') \ e^{ \ih H t} \ F^{\dag}_k (\l_2') \right)
&(3.11)
\cr }
$$
where the trace is now over the Hilbert space of the component
system.

The last part of the integrand may be written
$$
\eqalignno{
\prod_{k=1}^N \ \Tr \left(
F_k (\l_2 ) \ e^{ -\ih H t } 
\right. & \left.
\ F_k (\l_1) \ \rho
\ F^{\dag}_k (\l_1') \ e^{ \ih H t} \ F^{\dag}_k (\l_2') \right)
\cr
&= \left[ \Tr \left(
F_k (\l_2 ) \ e^{ -\ih H t } \ F_k (\l_1) \ \rho
\ F^{\dag}_k (\l_1') \ e^{ \ih H t} \ F^{\dag}_k (\l_2') \right)
\right]^N
\cr
&= \left[
e^{ i ( \l_1 + \l_2 - \l_1' - \l_2' ) } p(y,y)
+ e^{ i ( \l_2 - \l_2' ) } p(n,y)
\right.
\cr
& \quad \quad
\left.
+ e^{ i ( \l_1 - \l_1' ) } p (y,n) + p (n,n)
\right.
\cr
& \quad \quad \quad \quad
\left.
+ e^{ i ( \l_1 + \l_2 - \l_2' ) } D ( y,y | n,y )
+ e^{ i ( \l_2 - \l_1' - \l_2' ) } D ( n,y | y,y )
\right.
\cr
& \quad \quad \quad \quad \quad \quad
\left.
+ e^{ i \l_1 } D ( y,n | n,n )
+ e^{ -i \l_1' } D (n,n | y,n ) \right]^N
&(3.12)
\cr }
$$
where $ p(y,y) $, $D(y,y|n,y) $ {\it etc.} denote the
probabilities and decoherence functionals for each component system,
that is
$$
\eqalignno{
p(y,y) &=
\Tr \left( P \ e^{ - \ih H t } \ P \ \rho \ P \ e^{ \ih H t} 
\right)
&(3.13)
\cr
p(n,y) &=
\Tr \left( P \ e^{ - \ih H t } \ \P \ \rho \ \P \ e^{ \ih H t} 
\right)
&(3.14)
\cr
p(y,n) &=
\Tr \left( \P \ e^{ - \ih H t } \ P \ \rho \ P \ e^{ \ih H t} 
\right)
&(3.15)
\cr
p(n,n) &=
\Tr \left( \P \ e^{ - \ih H t } \ \P \ \rho \ \P \ e^{ \ih H t} 
\right)
&(3.16)
\cr
D(y,y|n,y) &=
\Tr \left( P \ e^{ - \ih H t } \ P \ \rho \ \P \ e^{ \ih H t} 
\right)
&(3.17)
\cr
D(n,y|y,y) &=
\Tr \left( P \ e^{ - \ih H t } \ \P \ \rho \ P \ e^{ \ih H t} 
\right)
&(3.18)
\cr
D(y,n|n,n) &=
\Tr \left( \P \ e^{ - \ih H t } \ P \ \rho \ \P \ e^{ \ih H t} 
\right)
&(3.19)
\cr
D(n,n|y,n) &=
\Tr \left( \P \ e^{ - \ih H t } \ \P \ \rho \ P \ e^{ \ih H t} 
\right)
&(3.20)
\cr }
$$ 
The probabilities obey the simple relation
$$
p(y,y) + p(n,y) + p(y,n) + p(n,n) = 1
\eqno(3.21)
$$
The off-diagonal terms of the decoherence functional satisfy
$$
\eqalignno{
D(y,y|n,y) + D(y,n|n,n) &= 0
&(3.22)
\cr
D(n,y|y,y) + D(n,n|y,n) &= 0
&(3.23)
\cr }
$$
The latter relations are particular to decoherence functionals
of histories characterized by projections at two moments of time.
They are consistent with the general property that the sum of
off-diagonal terms is zero. Also, since the second relation is the
complex conjugate of the first, they imply that all the off-diagonal
terms of the decoherence functional may be recovered from a single
complex number, which may be taken, for example, to be 
$D(y,y|n,y)$.

\subhead{\bf 3(C). Evaluation for Large $N$}

The integral (3.11), with (3.12) inserted may be evaluate exactly by
muliple use of the binomial expansion. This is carried out in
Appendix A. More useful is an approximate evaluation for large $N$.
To do this, note that
$$
\eqalignno{
\Bigl| \Tr \left(
F_k (\l_2 ) \ e^{ -\ih H_k t } \ F_k (\l_1) 
\right. & \left.
\ \rho_k
\ F^{\dag}_k (\l_1') \ e^{ \ih H_k t} \ F^{\dag}_k (\l_2') \right)
\Bigr|^2 
\cr
& \le
\Tr \left(
F_k (\l_2 ) \ e^{ -\ih H_k t } \ F_k (\l_1) \ \rho_k
\ F^{\dag}_k (\l_1) \ e^{ \ih H_k t} \ F^{\dag}_k (\l_2) \right)
\cr
& \quad \times
\Tr \left(
F_k (\l_2' ) \ e^{ -\ih H_k t } \ F_k (\l_1') \ \rho_k
\ F^{\dag}_k (\l_1') \ e^{ \ih H_k t} \ F^{\dag}_k (\l_2') \right)
\cr
& = 1
&(3.24)
\cr}
$$
The first relation, the inequality, is an elementary 
generalization of the inequality
$$
\bigl| D(\a, \a') \bigr|^2 \ \le \ D(\a,\a) \ D(\a',\a')
\eqno(3.25)
$$
obeyed by the decoherence functional [\cite{DoH}]. The second relation,
equality with unity, follows from the fact that $F_k(\l)$ obeys the
relation
$$
F_k(\l) F^{\dag}_k (\l) = 1
\eqno(3.26)
$$
as is easily shown.

The norm of the term raised to the power $N$ when Eq.(3.12) is
inserted in the integral (3.11) is
therefore less than or equal to $1 $, with equality if 
and only if all the $\l$'s
are zero. It follows that for very large $N$, the integral over the
$\l$'s in Eq.(3.11) 
is dominated by values of the $\l$'s close to zero, and the
integral may therefore be evaluated by exanding about this point.
(Note that we could not have used the inequality (3.24) if we had made
the simplification  $ \left( P_{\n_2} \right)^2 = P_{\n_2} $ in the
final projection).

We now make the following change of variables:
$$
\mu_1 = { \l_1 + \l_1' \over 2}, \quad
\xi_1 = {\l_1 - \l_1' \over 2}, \quad
\mu_2 = { \l_2 + \l_2' \over 2}, \quad
\xi_2 = {\l_2 - \l_2' \over 2}.
\eqno(3.27)
$$
The integral expression for the decoherence functional
then becomes,
$$
\eqalignno{
D(\n_1, \n_2 | \n_1', \n_2 ) 
&= {1 \over (2 \pi)^4 }
\ \int d \mu_1 \ d \xi_1 \ d \mu_2 \ d \xi_2
\ e^{ i \mu_1 (\n_1' - \n_1) -i \xi_1 (\n_1 +  \n_1')
- 2 i \xi_2 \n_2}
\cr
& \times
\left[ e^{2 i (\xi_1 + \xi_2) } p(y,y) 
+ e^{2i \xi_2} p(n,y) + e^{2i \xi_1} p(y,n) + p(n,n)
\right.
\cr & \quad \quad
\left.
+ e^{ i (\mu_1 + \xi_1 + 2 \xi_2) } D(y,y|n,y)
+ e^{i  (- \mu_1 + \xi_1 + 2 \xi_2 ) } D(n,y|y,y)
\right.
\cr & \quad \quad \quad \quad
\left.
+ e^{ i ( \mu_1 + \xi_1 ) } D(y,n|n,n)
+ e^{ i ( - \mu_1 + \xi_1 ) } D(n,n|y,n) \right]^N
&(3.28)
\cr }
$$ 
Since the integrand is independent of $\mu_2$, the integral over
$\mu_2$ may be carried out. As explained above, for very large $N$,
the integral over the remaining variables
may then be carried out by expanding the integrand about
$ \mu_1 = \xi_1 = \xi_2 = 0 $. To quadratic order, the term
in brackets raised to the power $N$ has the form,
$$
\left[ 1 + i {\bf x} \cdot {\bf v} 
- {\bf x}^T M {\bf x} \right]^N
\eqno(3.29)
$$
where $ {\bf x} $ is the three-vector $ (\mu_1, \xi_1, \xi_2 )$,
${\bf v}$ is a three-vector, and $M$ is a symmetric
$ 3 \times 3 $ matrix.
The components of ${\bf v}$ may be read off from the following:
$$
\eqalignno{
{\bf x} \cdot {\bf v}
=& 2 \left( p(y,y) + p(y,n) \right) \xi_1
\cr
& \quad + 2 \left( p(y,y) + p(n,y) + D(y,y|n,y) + D(n,y|y,y) \right) \xi_2
\cr
=& 2 p_0 \xi_1 + 2 p_t \xi_2
&(3.30)
\cr }
$$
where the properties of the decoherence functional (3.22), (3.23)
have been used, 
and $p_0$ and $p_t$ denote the probabilities of the
alternative ``yes'' at times $0$ and $t$ respectively, {\it i.e}, 
$$
p_0 = \Tr \left( P \rho \right), \quad\quad
p_t = \Tr \left( P e^{ \ih H t} \rho e^{ - \ih H t} \right)
\eqno(3.31)
$$
We also denote the probabilites of the ``no'' alternatives
at times $0$ and $t$ by ${\bar p}_0$ and ${\bar p}_t$
respectively.
Hence we have $p_0 + {\bar p}_0 = 1$, and
$ p_t + { \bar p}_t =1 $.
Similarly, the components of $M$ may be read off from
$$
\eqalignno{
{ \bf x}^T M {\bf x} =&
2 p_0 \xi_1^2 
+ 2 \left( 2 p(y,y) + D(y,y|n,y) + D(n,y|y,y) \right) \xi_1 \xi_2
\cr &
+ 2 p_t \xi_2^2
+ 2 \left( D(y,y|n,y) - D(n,y|y,y) \right) \mu_1 \xi_2 
&(3.32)
\cr }
$$
Again the properties (3.22), (3.23) have been used.

For large $N$,
$$
( 1 + z )^N \approx e^{ N ( z - \half z^2 ) }
\eqno(3.33)
$$
hence the decoherence functional now has the form
$$
\eqalignno{
D(\n_1, \n_2 | \n_1', \n_2 ) 
= & \int  d \mu_1 d \xi_1 d \xi_2
\ \exp \left(
i \mu_1 ( \n_1 - \n_1')
\right.
\cr & \left.
- i \xi_1 ( \n_1 + \n_1' - 2 N p_0 )
- i \xi_2 ( 2 \n_2 - 2 N p_t ) 
- {\bf x}^T A {\bf x} \right)
&(3.34)
\cr }
$$
where here, and in what follows, we will drop overall normalization
factors (these are readily recovered if required).
$A$ is a symmetric $3 \times 3 $ matrix, defined by
$$
{\bf x}^T A {\bf x}
= N {\bf x}^T M {\bf x} - \half N ( {\bf x} \cdot {\bf v} )^2
\eqno(3.35)
$$
Its explicit components, which will be important below, are
$$
\eqalignno{
A_{00} & = 0, \quad A_{01} = 0
&(3.36)
\cr
A_{02} & = i N \ {\rm Im} \left( D(y,y|n,y) \right)
&(3.37)
\cr
A_{11} & = 2 N p_0 {\bar p}_0
&(3.38)
\cr
A_{12} & = 2 N \left[ p(y,y) - p_0 p_t + {\rm Re} 
\left( D(y,y|n,y) \right) \right]
&(3.39)
\cr
A_{22} & = 2 N p_t {\bar p}_t
&(3.40)
\cr }
$$
The integrals may then be carried out, with the result,
$$
\eqalignno{
D(\n_1, \n_2 | \n_1', \n_2 ) 
= \exp & \left( - \a (\n_1 - \n_1')^2
- \b (\n_1 + \n_1' - 2 N p(y_0) )^2
\right.
\cr
\quad & \left. - i \gamma ( \n_1 - \n_1') (  \n_2 - N p_t )
\right.
\cr
\quad & \left. - i \nu ( \n_1 - \n_1') 
( \n_1 + \n_1' - 2 N p_0
\right)
&(3.41)
\cr }
$$
where the real coefficients $\a, \b, \gamma, \nu $
are given by
$$
\eqalignno{
\a = & { A_{11} A_{22} - A^2_{12} \over 4
A_{11} ( i A_{02} )^2 }
&(3.42)
\cr
\b = & { 1 \over 4 A_{11} }
&(3.43)
\cr
\gamma = & { 1 \over i A_{02} }
&(3.44)
\cr
\nu = & { A_{12} \over 2 A_{11} ( i A_{02} ) }
&(3.45)
\cr }
$$
This is now the decoherence functional for precisely specified
values of $\n$. 

To complete the calcuation of the decoherence
functional a further coarse-graining over $\n$ is required,
corresponding to imprecise specification of $\n$. This involves
summing $\n_1$, $\n_1'$ and $\n_2$ over ranges of
values, and is most easily achieved, at least approximately,
by taking $\n$ to be continuous and integrating with a
Gaussian smearing. Denoting the coarse-grained
values of $\n$ by $\N$, we have
$$
\eqalignno{
D(\N_1, \N_2 | \N_1', \N_2 ) & =
\int d \n_1 \ d \n_1' \ d \n_2 
\ D( \n_1, \n_2 | \n_1', \n_2 )
\cr
\times &
{ 1 \over (2 \pi \sigma^2 )^{3/2} } 
\ \exp \left( 
- { (\n_1 - \N_1)^2 \over 2 \sigma^2 }
- { (\n_1' - \N_1')^2 \over 2 \sigma^2 }
- { (\n_2 - \N_2)^2 \over 2 \sigma^2 }
\right)
&(3.46)
\cr}
$$
where the ranges of integration are $ - \infty $ to $+ \infty $.
The coarse-grained variables $\N$ are understood to have
significance only to up order $\sigma$.

The integrals are most easily carried out by changing variables
to $\n_1 + \n_1'$ and $\n_1 -\n_1'$, and by making use of the
formula
$$
\eqalignno{
\int dx & \ \exp \left( 
- a (x -x_1)^2 - b (x- x_2)^2 + i c x \right) 
\cr 
= & \left( { \pi \over (a+b) } \right)^{\half}
\ \exp \left(
- { c^2 \over 4 (a + b) }
- { ab \over (a + b ) } ( x_1 - x_2 )^2
+ i { c \over ( a+b) } ( a x_1 + b x_2 )  
\right)
&(3.47)
\cr }
$$
One thus obtains
$$
\eqalignno{
D( \N_1, \N_2  | \N_1', & \N_2) 
\cr
= \exp \left( 
\right. & \left.
- \tilde \a ( \N_1 - \N_1')^2
- \tilde \b (\N_1 + \N_1' - 2 N p_0 )^2
\right. \cr & \left.
- \tilde \epsilon (\N_2 - Np_t)^2
- \tilde \phi (\N_2 - N p_t ) ( \N_1 + \N_1' - 2 N p_0 )
\right. \cr & \left.
- i \tilde \gamma ( \N_1 - \N_1') ( \N_2 - N p_t )
- i \tilde \nu ( \N_1 - \N_1' ) ( \N_1 + \N_1' - 2 N p_0 )
\right)
&(3.48)
\cr }
$$
where
$$
\eqalignno{
\tilde \a &= { b \over ( 1 + 4 \s^2 b )},
\quad b = \a + { \gamma^2 \s^2 \over 2} 
+ { \nu^2 \s^2 \over ( 1 + 4 \s^2 \b) }
&(3.49)
\cr
\tilde \b &= { \b \over ( 1 + 4 \s^2 \b) }
+ { \s^2 \nu^2 \over ( 1 + 4 \s^2 b) ( 1 + 4 \s^2 \b)^2 }
&(3.50)
\cr
\tilde \epsilon &= { \gamma^2 \s^2 \over ( 1 + 4 \s^2 b ) }
&(3.51)
\cr
\tilde \phi & = { 2 \s^2 \nu \gamma \over ( 1 + 4 \s^2 b )
( 1 + 4 \s^2 \b) }
&(3.52)
\cr
\tilde \gamma & = { \gamma \over ( 1 + 4 \s^2 b )}
&(3.53)
\cr
\tilde \nu & = { \nu \over ( 1 + 4 \s^2 b ) ( 1 + 4 \s^2 \b ) }
&(3.54)
\cr}
$$
Eq.(3.48) is the decoherence functional
for densities of the ``yes'' alternative, coarse-grained to a width
$\s$, and for histories characterized by projections at two moments
of time.

\subhead{\bf 3(D). The Degree of Decoherence}

A reasonable measure of approximate decoherence is the
size of the off-diagonal terms in comparison to the probabilities,
Eq.(1.8). Denoting the diagonal parts of the decoherence functional by
$ p(\N_1, \N_2)$, we have,
$$
{ \bigl| D(\N_1, \N_2 | \N_1', \N_2 ) \bigr|
\over p(\N_1, \N_2)^{\half} p(\N_1', \N_2)^{\half} }
= \exp \left( - ( \tilde \a - \tilde \b) ( \N_1 - \N_1')^2 \right)
\eqno(3.55)
$$
hence the degree of approximate decoherence is controled by
$\tilde \a - \tilde \b$. Since the coarse-grained density $\N$ has
significance only up to order $\s$, the degree of decoherence, which
we denote $\epsilon$, is at worst given by
$$
\epsilon = \exp \left( - (\tilde \a - \tilde \b) \s^2 \right)
\eqno(3.56)
$$

For small $\s$, $\tilde \a \approx \a $ and 
$ \tilde \b \approx \b $.
We will see in the next section that, for the models we are
interested in, the probabilities $p_0$, $p_t$, {\etc} 
for the component systems
are of order $1$, whilst the off-diagonal
terms of the decoherence functional are much smaller. 
>From Eqs.(3.46)--(3.40),
this means
that $ \a >> \b $, the important term in $\a$ is the term 
$A_{02}$, and the terms $A_{11}$, $A_{22}$ and $A_{12}$ are
important only in their $N$-dependence. It is also convenient to
write the coarse-graining parameter $\s$ as a fraction $f$ of the
total particle number $N$, so $\s = f N$.
The degree of decoherence is therefore of the form
$$
\epsilon \approx \exp \left(
- { N \Gamma 
f^2 \over \left( {\rm Im} D(y,y|n,y) \right)^2 } \right)
\eqno(3.57)
$$
where
$$
\Gamma = p_t {\bar p}_t - { \left( p(y,y) -p_0 p_t + {\rm Re}
D(y,y|n,y) \right)^2 \over p_0 {\bar p}_0 }
\eqno(3.58)
$$

This is the expected result, and the main technical result of this
paper.  The degree of decoherence improves with increasing $N$. It
also improves as the degree of decoherence of the component systems
improves ({\it i.e.}, as $| D(y,y|n,y)| $ gets smaller). Moreover, 
decoherence also relies on the factor $f$ not being too small. This
means that the density $\n$ must be partitioned into macroscopically
distinct sets for there to be sufficient decoherence.

Note that Eq.(3.57) invites one to use the ratio of
$ ({\rm Im} D (y,y|n,y) )^2 $ to $\Gamma $ as a measure
of approximate decoherence of the component systems (rather than,
for example, the condition (1.8)) -- not an immediatey obvious
measure. Having said that, it will in fact turn out that for the
model of the next section, the probabilities are of order $1$,
so $\Gamma$ is of order $1$.  (See figure 3.)

Note also that the decoherence functional (3.48)  involves only the
imaginary part of the decoherence functional of the component
systems. As we shall see in the next section, it is possible to have
$ {\rm Re} D(y,y|n,y) = 0 $ but $ {\rm Im} D(y,y|n,y) \ne  0 $.
This means that the component systems could be exactly consistent
but the total system not exactly consistent. There is no
contradiction since the decoherence functional for the total
system is a sum of products of the decoherence functionals for the
component systems, so $ {\rm Re} D(y,y|n,y) = 0$ does not imply that
$ {\rm Re} D(\N_1, \N_2 | \N_1', \N_2 ) $ is exactly diagonal. Exact
{\it decoherence} of the component systems (rather than just
consistency), however, does imply exact decoherence of the total
system. The significance of this, if anything, is to underscore
decoherence, Eq.(1.7), as a physically more meaningful condition
than consistency, Eq.(1.6).

Given decoherence we may now examine the probabilities for the
histories. These are given by
$$
\eqalignno{
p( \N_1, \N_2)
=  \exp & \left( - 4 \tilde  \b (\N_1 - N p_0 )^2 
- \tilde \epsilon (\N_2 - Np_t)^2
\right.
\cr
& \quad \quad 
\left. 
- 2 \tilde \phi (\N_2 - N p_t ) ( \N_1 - N p_0 ) \right)
&(3.59)
\cr }
$$
This means that the probabilities for $\N_1$ only, or
for $\N_2$ only, are peaked about $Np_0$ or $Np_t$, as one might
expect. Beyond this, the form of the probabilitiy distribution
(3.59) is not of much significance. The model is too simple for us
to expect approximate evolution equations for $\N$.

There is nothing in this section that refers to the notion
of conservation.  The above result is essentially a combinatoric
one. As we will see in the next section, the role of conservation is
to ensure that $ | D(y,y|n,y) | $ is small, and moreover,
becomes smaller as the coarse-graining volume increases.

\head{\bf 4. The Spin System Model}

We now describe a particular model which is of the type discussed in
Section III, and in which we expect decoherence through approximate
conservation.

\subhead{\bf 4(A). The Model}

The model consists of a chain of very large number $M$ of locally coupled
``atoms'', each of which can be in one of two states, call them spin
up and spin down. (We follow Ref.[\cite{Fey}]).
The spins interact via the Hamiltonian
$$
H = - {\chi \over 2} \sum_{n=1}^M \ {\overrightarrow \s}_n 
\cdot {\overrightarrow \s}_{n+1}
\eqno(4.1)
$$
where ${\overrightarrow \s}_n$ is the 3-vector whose components are the Pauli
matrices. We impose periodic boundary conditions, 
${\overrightarrow \s}_{N+1} = {\overrightarrow \s}_1$.
Up to an additive constant, the Hamiltonian may be written
$$
H = - \chi \sum_{n=1}^M \ p^{n,n+1}
\eqno(4.2)
$$
Here, $p^{n,n+1}$ is a spin exchange operator. It has the effect of
leaving aligned spins alone, and interchanging oppositely aligned
spins. For example,
$$
p^{1,2} \ | \up \do \ra = | \do \up \ra
\eqno(4.3)
$$

The ground state is the state with all spins pointing up and has
eigenvalue zero.
The next excited states consist of the $M$
states for which one spin is down and the rest up, and we denote them
$ | \phi_k \ra $, where
$$
\eqalignno{
| \phi_1 \ra &= | \do \up \up \cdots \up \ra
&(4.4)
\cr
| \phi_2 \ra &= | \up \do \up \cdots \up \ra
&(4.5)
\cr }
$$
and so on, {\it i.e.}, $ | \phi_k \ra $ denotes the state in which
spin number $k$ is down and the rest are up. These are the so-called
spin waves [\cite{Fey}].
Then it is readily seen that
$$
\eqalignno{
p^{k,k+1} \ | \phi_k \ra &= | \phi_{k+1} \ra
&(4.6)
\cr
p^{k,k+1} \ | \phi_{k+1} \ra & = | \phi_k \ra
&(4.7)
\cr}
$$
and
$$
p^{k,k+1} \ | \phi_n \ra = 0
\eqno(4.8)
$$
if $ n < k$ or $n > k+1 $.

The states $| \phi_k \ra $ are not eigenstates of the Hamiltonian,
but the combinations
$$
| \psi_{\ell} \ra = {1 \over \sqrt{M} } \sum_{k=1}^M 
\ \exp\left( { 2 \pi i \ell k \over M } \right) \ | \phi_k \ra
\eqno(4.9)
$$
are, with eigenvalues $E_{\ell} = - 2 \chi \ \cos \left( { 2 \pi \ell
\over M} \right) $, and $\ell = 1, \cdots M$. Their normalization
follows from the identity,
$$
{ 1 \over M } \sum_{k=1}^M \ \exp \left( { 2 \pi i k ( n - m ) \over
M} \right) = \delta_{nm}
\eqno(4.10)
$$
where $\delta_{nm}$ is the Kronecker delta.

\subhead{\bf 4(B). Coarse Grainings}

There are two conserved quantities for this model: the total energy
and the total spin. We will study coarse-grainings which ask
for the total spin in a region of the chain. We take the
crudest coarse-graining, which is to  divide the chain into two
regions, region 1 and region 2, of size $M_1$ and $M_2$, where $M_1
+ M_2 = M$. We will construct projections which ask for the total
number of down spins in region 1. 

The model so far is not yet exactly of the type described in Section
III because there, a non-interacting Hamiltonian was assumed, whereas
here, we have an interacting one. To get around this, and hence to
use the results of Section III, we do the following. First, we
restrict attention to the subspace in which there is a fixed number
$N$ of spins pointing down. This subspace is invariant under
Hamiltonian evolution, so the set of $N$ downward pointing spins may be
regarded as a set of $N$ interacting ``particles''.
Second, and more importantly,  we assume that $ 1 << N << M $. This
is a kind of ``dilute gas'' assumption, and we expect it to
allow us to neglect the interaction between downward
pointing spins. 

More precisely, we replace the $N$ particle
Hilbert space $ {\cal H}_N$ with a tensor product
of $N$ Hilbert spaces ${\cal H}_1$, where ${\cal H}_1$ denotes the
Hilbert space of states with one spin pointing down described above.
We also take the Hamiltonian $H_N$ for the $N$ particles to be of the form 
$$
\eqalignno{
H_N & =  H_1 \otimes 1 \otimes 1 \otimes \cdots
\cr \quad
& \quad +  1 \otimes H_1 \otimes 1 \otimes \cdots 
\cr \quad 
& \quad + \cdots 
&(4.11)
\cr }
$$
where $H_1$ is the Hamiltonian for the states with one spin pointing
down, operating on ${\cal H}_1$.
The Hamiltonian is thus of the form required for the results
of Section III to be applicable.

Clearly some work is required to fully justify this
approximation. The Hilbert space ${\cal H}_N$ does not have exactly
the same dimension is the tensor product of the $N$ spaces ${\cal
H}_1$, although they are close for sufficiently large $N$ and $M$.
Also, one would expect the effective Hamiltonian to include a term
in the Hamiltonian preventing two down spins from occupying the same
site, and a term describing the interacting between neighbouring
down spins. What we are assuming, in effect, is that these extra
terms can be neglected if the gas of downward pointing spins is
sufficiently dilute. 

Even if this approximation cannot be justified, we could always
postulate a model of the above type. This would admittedly be
unphysical, but at least it provides a framework in which we can
investigate the mathematical properties we are interested in, which
is the main aim of this paper.

The projection $P_{\n}$ onto the total spin in region $1$, is constructed
from the projection for the individual particles, as described in
Section III, Eq.(3.5). For each individual particle we introduce the
projections,
$$
\eqalignno{
P &= \sum_{k=1}^{M_1} \ | \phi_k \ra \la \phi_k |
&(4.12)
\cr
\P &= \sum_{k={M_1+1}}^M \ | \phi_k \ra \la \phi_k |
&(4.13)
\cr }
$$
In the $1$-particle subspace, $P$ and $\P$ ask whether the
downward spin is in region $1$ or not in region $1$ ({\it i.e.}, 
in region $2$), respectively. We also denote these alternatives $y$
and $n$, respectively, in accordance with the notation of Section III.
As shown in Section III, the computation of the decoherence
functional onto histories in which the total spin in region 1 is
specified reduces to a computation of the decoherence functional for
the component systems.

\subhead{\bf 4(C). The Decoherence Functional of the Component Systems}

The Hamiltonian for the 1-particle subspace ${\cal H}_1$
may be written, 
$$ 
H_1 = \sum_{\ell=1}^M \
E_{\ell} \ | \psi_{\ell} \ra \la \psi_{\ell} | 
\eqno(4.14)
$$ 
and hence the
unitary evolution operator is 
$$ 
e^{-iH_1t} = \sum_{\ell=1}^M \ e^{-i
t E_{\ell}}   \ | \psi_{\ell} \ra \la \psi_{\ell} | 
\eqno(4.15)
$$ 
The Hamiltonian may also be written, 
$$ 
H_1 = - \chi \sum_{k=1}^M \left( |
\phi_k \ra \la \phi_{k-1} | + | \phi_{k-1} \ra \la \phi_k | \right)
\eqno(4.16)
$$

The decoherence functional for a simple two-time history is
$$
D( \a_1, \a_2 | \a_1', \a_2 ) = {\rm Tr} \left( P_{\a_2} (t)
P_{\a_1} \rho P_{\a_1'} \right)
\eqno(4.17)
$$
where $P_{\a_2}(t) = e^{iH_1 t} P_{\a_2} e^{-iH_1 t} $, and $P_{\a}$
denotes $P$ and $\P$. Take an initial state which is a
superposition states in which the particles is in region $1$ and $2$,
$$
| \Psi \ra = {1 \over \sqrt{2} } \left( 
| \phi_{k_1} \ra + | \phi_{k_2} \ra \right)
\eqno(4.18)
$$
where $ 1 \le k_1 \le M_1 $, and $ M_1 +1 \le k_2 \le M $.
Then
$$
P_{\a}  \ | \Psi \ra \la \Psi | \ P_{\a'}
= \half \ | \phi_{k_{\a}} \ra \la \phi_{k_{\a'}} |
\eqno(4.19)
$$
where for the moment we let $ \a $ take the values $1,2$,
corresponding to $y,n$.
The decoherence functional then is
$$
D(\a_1, \a_2 | \a_1', \a_2 ) = 
\half \la \phi_{k_{\a_1}} |
P_{\a_2} (t) | \phi_{k_{\a_1'}} \ra
\eqno(4.20)
$$
For this, we then have, for example
$$
\eqalignno{
D(y,y|n,y) &= \half \la \phi_{k_1} | P (t) | \phi_{k_2} \ra
&(4.21)
\cr
p(n,y) &= \half \la \phi_{k_2} | P (t) | \phi_{k_2} \ra
&(4.22)
\cr}
$$
and so on.

Let $P_{\a_2} = P $. Then it is readily shown that
$$
\la \phi_n | P (t) | \phi_{n'} \ra  = {1 \over M} \sum_{\ell=1}^M 
\ \sum_{\ell'=1}^M \  e^{it ( E_{\ell} - E_{\ell'} ) }
\ \exp \left( { 2 \pi i ( n \ell - n' \ell' ) \over M} \right)
\ d( \ell, \ell')
\eqno(4.23)
$$
where
$$
d(\ell, \ell' ) = \sum_{k=1}^{M_1} \ \exp \left( - { 2 \pi i k ( \ell -
\ell') \over M } \right)
\eqno(4.24)
$$
Now, writing $x = 2 \pi ( \ell - \ell') / M $, we have
$$
d (\ell, \ell') = \sum_{k=1}^{M_1} \ e^{- i k x} = { ( e^{-iM_1 x} - 1) \over
( 1 - e^{ix} ) }
\eqno(4.25)
$$
The properties of the decoherence functional may be understood
through $d(\ell, \ell')$. Clearly, if $M_1 = M$, then $ d (\ell, \ell')
= M \delta_{\ell \ell'} $, and $P (t)$ would be diagonal in the
$|\phi_n \ra $'s. More generally, if $M_1$ and $M$ are very large,
then $d(\ell, \ell')$ is very small when $\ell \ne \ell'$, and
$ d(\ell, \ell) = M_1$. 

Curiously, when $M_1 = M_2$ the real part of the decoherence functional
vanishes.  Thus, this coarse-graining exhibits exact consistency, but not
exact decoherence.  It seems reasonable that this sort of consistency, where
the imaginary part of the decoherence functional is non-zero, will only
occur due to symmetries in the system and choice of histories.

\subhead{\bf 4(D). Numerical Results}

The decoherence functional for the component systems may be computed
numerically using Eq.(4.20). There are three parameters, $ t $, $M$ and
$M_1$. The decoherence functional for various ranges of the
parameters is computed below and the results plotted in figures 1--3.

\subhead{\bf 4(E). Summary of Results}

The results of the numerical calculation may be concisely summarized
as follows.
The probabilities $p(y,y)$, $p(y,n)$, {\it etc.} are of order $1$.
The off-diagonal terms of the
decoherence functional, divided by the probabilities, are typically
of order $ M^{- \half} $, hence the degree of decoherence for the
$1$-particle case is of this order. This degree of decoherence is
not particularly good (compared, for example, to the quantum
Brownian motion models, where one typically finds that the degree of
decoherence is an exponential function of the coarse-graining
parameters [\cite{DoH}]). It is no better than one would expect from the overlap
of two arbitrary vectors in a Hilbert space of dimension $M$ [\cite{Hil}]. 
Physically, this is not surprising, because the histories in
question differ by just one quantum of spin, and one would not
expect the interference between such spins to be substantially
supressed.

However, when inserted into the expression for the decoherence
functional of the $N$-particle case, (3.48), a vastly improved degree
of decoherence is obtained, for sufficiently large $N$.  Moreover,
the important feature about the decoherence functional (4.20) is that
the off-diagonal terms decrease as the size $M_1$ of the``smearing
volume" increases from zero (until $M_1$ reaches the value about
$M/2$). Correspondingly, the off-diagonal terms of the decoherence
functional of the $N$-particle case (3.48) decrease very rapidly as
$M_1$ increases from zero.

\head{\bf 5. Summary and Conclusions}

This paper marks a first step towards the problem of computing the
decoherence functional for hydrodynamic histories.
We considered systems of essentially non-interacting particles and
studied histories of projections onto densities of those particles.
Our principal aim was to show, in the context of some simple models,
how decoherence can come about as a result of approximate
conservation, in tune with the general ideas put forward by
Gell-Mann and Hartle [\cite{GH1,GH2}].

The results of Section IV show explicitly how approximate
decoherence is related to approximate conservation: the degree of
decoherence increases as the smearing volume increases (at least,
for volumes less than half the total volume of the system). The
degree of decoherence, however, is not very good. This is because we
considered single particle models in Section IV, and the histories
differ by just one quantum of spin. The histories are
therefore not ``macroscopically distinct'' and one would expect that
interference between them could still be quite noticeable.
Differently put, it is because the fluctuations in variables in
question ({\it e.g.}, number density) are comparable to the values
of the variables themselves.

To obtain adequate decoherence, it is necessary to have a large
number $N$ of particles and partition the number density by large
ranges $\s$. It is here that the results of Section III came in. We
found that when the histories do differ by macroscopically
significant amounts, interference is destroyed very efficiently.
Clearly what is happening here is that the fluctuations in the
variables are much smaller than the variables themselves.

>From this we conclude that, in these simple models, decoherence
requires {\it two} distinct phenomena: approximate conservation, and
large particle number partitioned into large ranges. Approximate
conservation ensures that the dynamical variables projected onto
become sufficiently slowly varying for sufficiently large smearing
volume. Large particle number partitioned into large ranges ensures
that the quantum fluctuations in the local densities are smaller than
the values of the variables themselves. These conclusions 
concure with the general expectations expressed in 
Refs.[\cite{GH1,GH2}].

The restriction to non-interacting particles may seem rather
unrealistic. It means that conservation may already be seen at the
one particle level, whilst for an interacting theory, it is only
seen for the whole collection of particles. Although physically
unrealistic, it had the mathematical advantage that the effects of
large particle number and approximate conservation could be cleanly
seperated. Moreover, one would not expect the inclusion of
interactions to substantially modify our conclusions (although this
is clearly an important extension to carry out). The point is that
here, it is the approximate conservation and large particle number
that produce decoherence, not the interactions. This is in stark
contrast to the vast majority of other models studied in the
literature on decoherence, in which it is interactions (usually with
another system) that are held responsible for decoherence.

This paper is, as stated, a first step in an extensive
investigation, and there are therefore many ways in which it may be
developed. Perhaps the next step is to seek a more sophisticated
model in which one would expect the probabilities for histories to
be peaked about interesting approximately deterministic evolution
equations. 

A technically very different but similar in spirit investigation is
that of Calzetta and Hu [\cite{CaH}], who considered the decoherence
of histories characterized by the values of $n$-point functions for
fields. They did not make any contact with the notion of
approximate conservation, but it is similar to our work in that it
is, to the best of our knowledge, the only other concrete
calculation of decoherence which avoids a system--environment split.
It would be of interest to find more detailed connections between
their work and ours.

These and other questions will be the topic of future publications.

\head{\bf Acknowledgements}

J.J.H. carried out part of this work at the Los Alamos National
Laboratory, and would like to thank James Anglin, Salman Habib and
Wojtek Zurek for useful conversations and hospitality. Part of this
work was also carried out at the University of Buenos Aires, where
the hospitality and useful converstaions of Mario Castagnino, Diego
Harari and Juan Pablo Paz were much appreciated. J.J.H. is
especially grateful to Jim Hartle for very many useful 
conversations over a long period of time, on decoherent histories
and particularly on hydrodynamic variables.
We are also very grateful to Ian Percival for useful conversations.

T.A.B. was supported by EPSRC.
J.J.H. was supported by a University Research Fellowship from the
Royal Society.

\head{\bf Appendix A: Exact Evaluation of the Decoherence Functional}

The integrals over the $\l$'s may be evaluated exactly, and
we present here the result of that calculation.
$$
\eqalignno{
D(\n_1, \n_2 | \n_1', \n_2 ) &= \Tr \left( P_{\n_2} \ e^{ - \ih H t} 
\ P_{\n_1} \ \rho \ P_{\n_1'} \ e^{ \ih H t} \right)
\cr
& = { 1 \over ( 2 \pi)^3 } 
\int_{-\pi}^{\pi} d \l_1
\int_{-\pi}^{\pi} d \l_1'
\int_{-\pi}^{\pi} d \l_2
\ e^{-2i \l_1 \n_1 + 2i \l_1' \n_1' - 2i \l_2 \n_2}
\cr
\quad \quad & \times  \left[
e^{ 2i ( \l_1 + \l_2 - \l_1' ) } p(y,y)
+ e^{ 2i \l_2 } p(n,y)
\right.
\cr
& \quad \quad
\left.
+ e^{ 2i ( \l_1 - \l_1' ) } p (y,n) + p (n,n)
\right.
\cr
& \quad \quad \quad \quad
\left.
+ e^{ 2i ( \l_1 + \l_2 ) } D ( y,y | n,y )
+ e^{ 2i ( \l_2 - \l_1') } D ( n,y | y,y )
\right.
\cr
& \quad \quad \quad \quad \quad \quad
\left.
+ e^{ 2i \l_1 } D ( y,n | n,n )
+ e^{ -2i \l_1' } D (n,n | y,n ) \right]^N
&(A1)
\cr }
$$

Now note that the term in brackets raised to the power $N$ may be
written 
$$
\left[ A + B e^{ 2 i \l_2} \right]^N
= \sum_{k=0}^N \left( { N \atop k} \right) \ A^{N-k} \ B^k
\ e^{2 i \l_2 k}
\eqno(A2)
$$
using the binomial expansion, where
$$
\eqalignno{
A &= e^{ 2i ( \l_1 - \l_1') } p(y,n) + p(n,n)
+ e^{ 2 i \l_1} D (y,n|n,n) + e^{-2i \l_1'} D(n,n|y,n)
&(A3)
\cr
B &= e^{ 2i (\l_1 - \l_1')} p(y,y) + p(n,y)
+ e^{ 2i \l_1} D(y,y|n,y) + e^{-2i \l_1'} D(n,y|y,y)
&(A4)
\cr }
$$
The integral over $\l_2$ may now be evaluated, with the result,
$$
D(\n_1, \n_2 | \n_1', \n_2 ) =
{ 1 \over (2 \pi)^2 } \int d \l_1 d \l_1'
\ e^{-2 i \l_1 \n_1 + 2 i \l_1' \n_1'}
\ \left( { N \atop \n_2} \right)
\ A^{N- \n_2} \ B^{\n_2}
\eqno(A5)
$$

We now use the binomial expansion again on $A^{N- \n_2}$
and $B^{\n_2}$. Write
$$
\eqalignno{
A &= e^{-2i \l_1'} C + D
&(A6)
\cr
B &= e^{-2i \l_1'} E + F
&(A7)
\cr }
$$
where
$$
\eqalignno{
C &= e^{2i \l_1} p(y,n) + D(n,n|y,n)
&(A8)
\cr
D & = p(n,n) + e^{2i \l_1} D(y,n|n,n)
&(A9)
\cr
E & = e^{2 i \l_1} p(y,y) + D(n,y |y,y)
&(A10)
\cr
F & = p(n,y) + e^{2 i \l_1} D(y,y|n,y)
&(A11)
\cr}
$$
Expanding $A^{N -\n_2} $ and $B^{\n_2}$,
the integral over $\l_1'$ may then be carried out, with 
the result,
$$
\eqalignno{
D(\n_1, \n_2 | \n_1', \n_2 ) =
{1 \over 2 \pi} \left( { N \atop \n_2} \right)
\int d \l_1 \ & e^{ -2i \l_1 \n_1}
\ \sum_{k=0}^{N-\n_2} \ \sum_{j=0}^{\n_2}
\ \left( { N - \n_2 \atop k } \right)
\ \left( { \n_2 \atop j} \right)
\cr
& \times
\ C^k \ D^{ N - \n_2 - k} \ E^j \ F^{\n_2 -j}
\ \delta_{k+j, \n_1'}
&(A12)
\cr}
$$

Repeating the binomial expansion of $C^k$ {\it etc.} on final time,
the last integration over $\l_1$ may be performed, and
we arrive at the result,
$$
\eqalignno{
D(\n_1, \n_2 | \n_1', \n_2 ) =
\left( { N \atop \n_2 } \right)
& \ \sum_{k=0}^{N-\n_2}
\ \sum_{j=0}^{\n_2}
\ \sum_{\ell =0}^k
\ \sum_{m=0}^{N-\n_2 -k}
\ \sum_{r=0}^j
\ \sum_{s=0}^{\n_2 -j}
\cr &\times
\ \left( { N - \n_2 \atop k} \right) \ \left( { \n_2 \atop j} \right)
\ \delta_{k+j, \n_1'} \ \delta_{ \ell + m + r + s, \n_1}
\cr &\times
\ \left( { k \atop \ell} \right)
\left[ D(n,n|y,n) \right]^{k-\ell} \ \left[ p(y,n) \right]^{\ell}
\cr &\times
\ \left( { N - \n_2 -k \atop m} \right)
\ \left[ p(n,n) \right]^{N- \n_2 - k - m} \ \left[ D(y,n|n,n) \right]^m
\cr &\times
\ \left( { j \atop r} \right)
\ \left[ D(n,y|y,y) \right]^{j-r} \ \left[ p(y,y) \right]^r
\cr &\times
\ \left( { \n_2 - j \atop s} \right)
\ \left[ p(n,y) \right]^{\n_2 - j -s } \ \left[ D(y,y|n,y) \right]^s
&(A13)
\cr }
$$

For reasonably small $N$, this expression can be evaluated
numerically. For large $N$ (and $\n_1$, $\n_1'$, $\n_2$), the sums
can be estimated by replacing them by integrals, and assuming that
the dominant contributions come from the middle of the ranges,
{\it i.e.}, from values of $j,k,\ell,m,r,s$ that are large
(although we have not carried this out explicitly).

\references
\def\pr{{\sl Phys.Rev.}}

\refis{CaH} E. Calzetta and B. L. Hu, in {\it Directions in General
Relativity}, edited by B. L. Hu and T. A. Jacobson (Cambridge
University Press, Cambridge, 1993).

\refis{CaL} A.Caldeira and A.Leggett, {\sl Physica} {\bf 121A}, 587 (1983).

\refis{DoH} H. F. Dowker and J. J. Halliwell, {\sl Phys. Rev.} {\bf
D46}, 1580 (1992).

\refis{Fey} R.Feynman, {\it Statistical Mechanics} (Addison-Wesley,
1972), Chapter 7.

\refis{FeV} R.Feynman and Vernon, {\sl Ann. Phys.} {\bf 24}, 118 (1963).

\refis{For} D. Forster, {\it Hydrodynamic Fluctuations, Broken
Symmetry and Correlation Functions} (Benjamin, Reading, MA, 1975).

\refis{GH1} M. Gell-Mann and J. B. Hartle, in {\it Complexity, Entropy 
and the Physics of Information, SFI Studies in the Sciences of Complexity},
Vol. VIII, W. Zurek (ed.) (Addison Wesley, Reading, 1990); and in
{\it Proceedings of the Third International Symposium on the Foundations of
Quantum Mechanics in the Light of New Technology}, S. Kobayashi, H. Ezawa,
Y. Murayama and S. Nomura (eds.) (Physical Society of Japan, Tokyo, 1990).

\refis{GH2} M. Gell-Mann and J. B. Hartle, {\sl Phys.Rev.} {\bf D47},
3345 (1993).

\refis{GH3} M. Gell-Mann and J. B. Hartle, Santa Barbara preprint
UCSBTH-95-28, gr-qc/9509054 (1995).

\refis{Gri} R. B. Griffiths, {\sl J.Stat.Phys.} {\bf 36}, 219 (1984);
{\sl Phys.Rev.Lett.} {\bf 70}, 2201 (1993).

\refis{Hal4} J. J. Halliwell, in {\it General Relativity and Gravitation
1992}, edited by R. J. Gleiser, C. N. Kozameh and O. M. Moreschi
(IOP Publishers, Bristol, 1993).

\refis{HaZ} J.J. Halliwell and A. Zoupas, to appear Phys. Rev. D (1995).

\refis{Har1} J. B. Hartle, in {\it Quantum Cosmology and Baby
Universes}, S. Coleman, J. Hartle, T. Piran and S. Weinberg (eds.)
(World Scientific, Singapore, 1991).  {\it The Quantum Mechanics of
Cosmology.}

\refis{Har6} J. B. Hartle, Santa Barbara preprint (1994), to appear
in {\it Proceedings of the Lanczos Centenary Meeting}.

\refis{HLM} J. B. Hartle, R. Laflamme and D. Marolf,
  \pr {\bf D51}, 7007 (1995).

\refis{Hil} W. Wootters, {\sl Phys.Rev.} {\bf D23}, 357 (1981).

\refis{JoZ} E.Joos and H.D.Zeh, {\sl Z.Phys.} {\bf B59}, 223 (1985).

\refis{Omn} R. Omn\`es, {\sl J.Stat.Phys.} {\bf 53}, 893 (1988).

\refis{Omn2} R.Omn`es, {\sl J.Stat.Phys.} {\bf 53}, 933 (1988);
{\bf 53}, 957 (1988);
{\bf 57}, 357 (1989);
{\sl Ann.Phys.} {\bf 201}, 354 (1990);
{\sl Rev.Mod.Phys.} {\bf 64}, 339 (1992).

\refis{PZ} J. P. Paz and W. H. Zurek, \pr {\bf D48}, 2728 (1993).

\refis{Zur} W. Zurek, in {\it Physical Origins of Time Asymmetry}, edited by 
J. J. Halliwell, J. Perez-Mercader and W. Zurek (Cambridge
University Press, Cambridge, 1994).

\endreferences

\eject\vfil

Figure 1.  The probabilities $p(1,1)$, $p(1,2)$, $p(2,1)$, $p(2,2)$ as a
function of the coarse-graining $M_1$.  This plot was produced at $t=1000$
for a string of $M=1000$ spins; the energy constant is $\chi = 1$.  Note
that for $M_1$ close to 500 (half the spins), the probabilities are
all close to 0.25.

\vfil

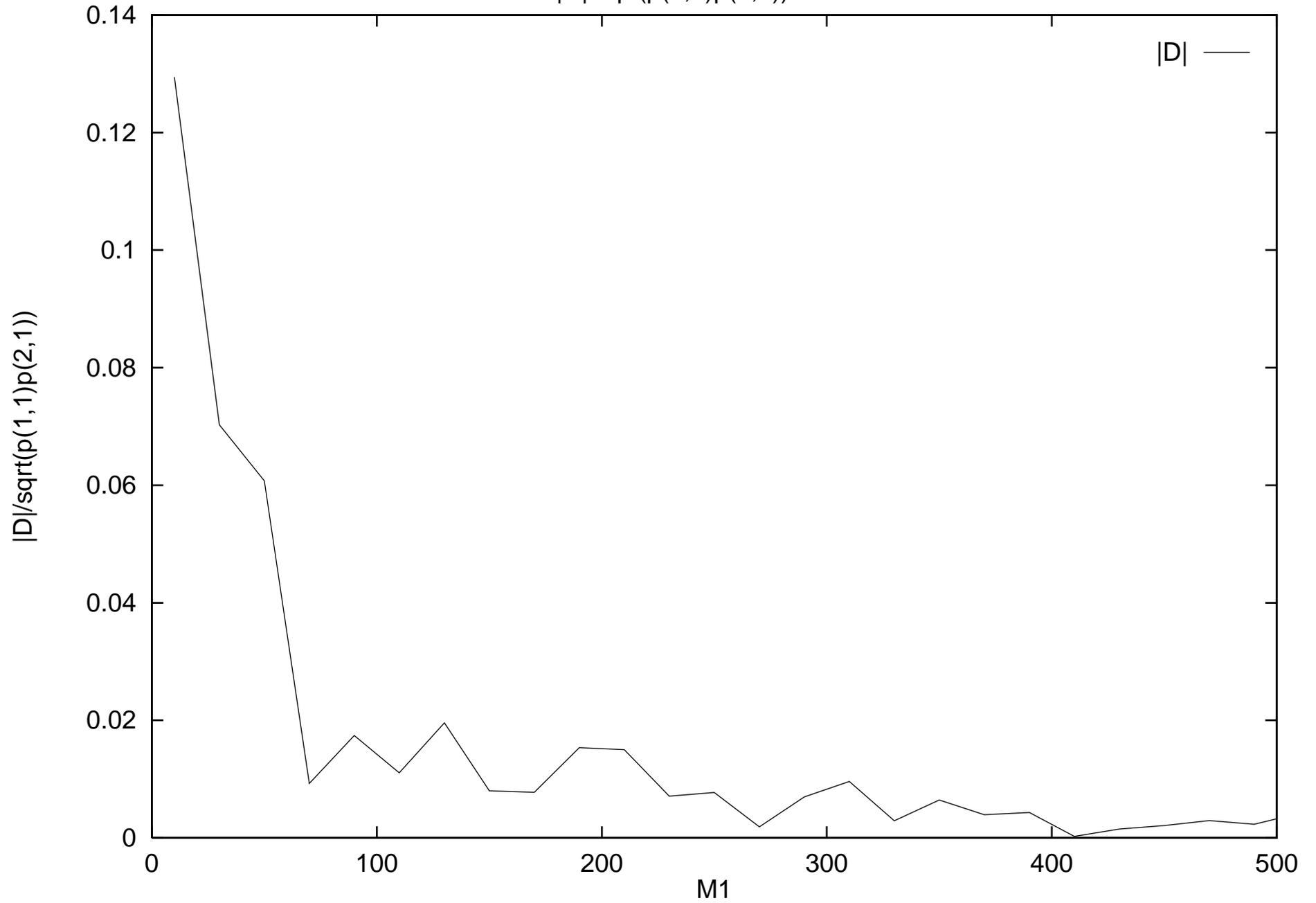
Figure 2.  $|D[1,1; 2,1]|/\sqrt{p(1,1) p(2,1)}$ vs. $M_1$.  For this plot
the parameters are $t=1000$, $M=1000$, $\chi=1$.  Note this quantitity is
highly peaked for low values of $M_1$, indicating that better decoherence
results from coarser graining.

\vfil

Figure 3.  $|{\rm Im} D[1,1; 2,1]|^2/\Gamma$ vs. $M_1$.  For this plot the
parameters are $t=1000$, $M=1000$, $\chi=1$.  This also exhibits peaking
for low values of $M_1$, indicating good decoherence for coarse graining.
Note also that the absolute value of this quantity is low; this will
produce very good decoherence for many-spin systems, as shown in equation
(3.57).

\vfil\end